# Development of an Extrudable Paste to build Mycelium-bound Composites

*Eugene Soh[‡,†], Zhi Yong Chew[‡,†], Nazanin Saeidi,[§,+], Alireza Javadian[§,+], Dirk Hebel[§,+], Hortense Le Ferrand[*,†,||]*

ABSTRACT.

Mycelium-bound composites are promising materials for sustainable packaging, insulation, fashion, and architecture. However, moulding is the main fabrication process explored to date, strongly limiting the ability to design the complex shapes that could widen the range of applications. Extrusion is a facile and low energy-cost process that has not been explored yet for mycelium-bound composites with design freedom and structural properties. In this study, we combine cheap, easily and commonly available agricultural waste materials, bamboo microfibres, chitosan, and mycelium from *Ganoderma Lucidum,* to establish a composite mixture that is workable, extrudable and buildable. We study the impact of bamboo fibre size, chitosan concentration, pH and weight ratio of bamboo to chitosan to determine the optimum growth condition for the mycelium as well as highest mechanical stiffness. The resulting materials have thus low energy costs, are sustainable and can be shaped in diverse forms easily. The developed composition is promising to further explore the use of mycelium-bound materials for structural applications using agricultural waste.

INTRODUCTION.

Mycelium-bound materials are composites made using a living organism, a fungus, growing onto a nutrient-rich substrate, typically plant-based such as straw, sawdust, cotton, etc. The mycelium is a hydrophobic branched network developed by the fungus to colonize the substrate, degrade it, and digest its nutrients.[1] Using mycelium as a living binder for biocomposites can be used to generate interesting properties in low-cost materials.[2,3] Indeed, mycelium-bound composite foams[4] or pressed laminates[5] were found to display strength[6–8] and antifire[8] and acoustic properties.[9,10] Mycelium-bound composites made from fungal species such as *Schizophyllum Commune*, *Pleurotus Ostreatus* or *Ganoderma Lucidum*, have found applications in packaging,[11] fashion,[12] thermal and acoustic insulation,[13,14] and architecture.[15–17] Till now, moulding is the method used to fabricate those materials:[18,19] the substrate and mycelium spawn are mixed together, poured into a mould until the mycelium growth is stopped by heating. To explore other applications and design possibilities, the fabrication technique should exhibit higher degrees of freedom in terms of shaping.

To obtain a high degree of freedom in designing and structuring materials, 3D printing is often a method of choice. 3D printing technologies have been applied to all material classes to create shapes and structures ranging from the nanometric to the macroscopic scales and for a large span of applications, including architecture.[20,21] Mycelium 3D printing has been explored by artists and designers such as Eric Klarenbeek, Chester Dols, the Blast studio, or the Offina Corpuscoli and Co-de-IT but the procedures and properties are yet to be studied systematically. Although the sustainability and affordability of 3D printing are questionable,[22] direct writing or robocasting is a simple method to build 3D structures using natural materials such as clay.[23] Other methods for 3D printing of organic composite materials comprise stereolithography where UV light cures the organic matrix, fused deposition modelling where heat is applied to a filament, inkjet printing of liquid droplets, lamination of sheets and laser beam processes.[24] In the case of living organic composites, the use of heat and of curing additives is compromised. Also, as compared to other manufacturing processes, extrusion is one of the least energy-demanding and therefore promising for low-cost and low-energy production of mycelium-bound materials (**Figure 1A**). Furthermore, mycelium-bound composites can exhibit mechanical properties similar to commercial Styrofoam, polyethylene (PE) and polyurethane (PU) foams, wool, up to low strength particle boards, at a much lower cost (**Figure 1B**). There is thus a large benefit in replacing synthetic foams or boards with extruded mycelium-bound materials.

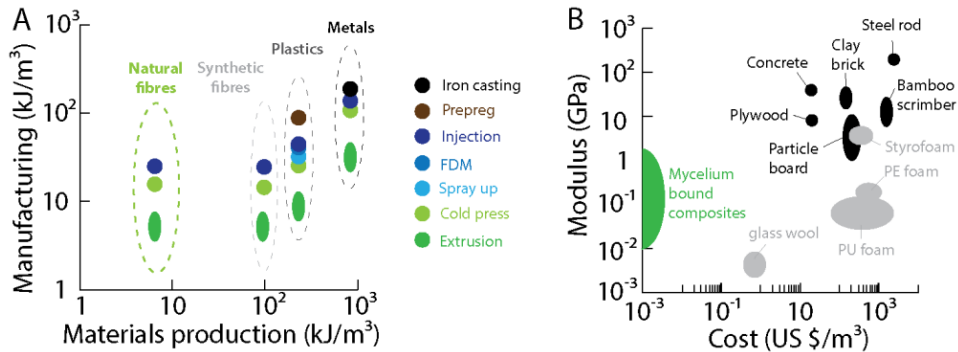

**Figure 1. Comparison of processes and materials in terms of energy cost, price, and stiffness. (A)** Energy intensity required by manufacturing processes as a function of the energy intensity required to produce the raw materials. The dashed areas highlight the raw materials considered, namely metals, plastics, synthetic and natural fibres. FDM stands for fused deposition modelling. Data extracted from refs.[25,26] **(B)** Young's modulus as a function of the cost per volume of material for mycelium-bound composites and other common materials. Black represents structural materials, grey thermally insulating foams, and green, mycelium-bound composites. Data extracted from standard commercial materials.

To be able to extrude mycelium-bound composites, it is necessary to develop an initial paste or mixture that has shape-retention properties. In the following, these two characteristics are termed workability and buildability.[27–29] Common strategies to increase workability and buildability use additives such as fibres and binders. For example, cementitious pastes could be extruded by adding microcellulose fibres,[27] adjusting the water content,[30] or tuning the viscosity with modifiers[28]. For mycelium materials, dry substrate-based structures without viscosity or submerged cultures with high fluidity have been reported,[31,32] both lacking the buildability character. Also, workability has been investigated on mycoprotein pastes, but for food application.[33] Cellulose nanofibrils have also been added to mycelium-wood mixtures using paddle mixing but the final composite was prepared by cold pressing to form a board.[34]

In this paper, we explore the use of bamboo chopped fibres and chitosan as ingredients to increase the workability and buildability of mycelium mixtures to build mycelium-bound composites using extrusion. The bamboo fibres will provide the nutrients for the growth of mycelium,[35,36] chitosan will serve as a rheological modifier to increase the workability and allow extrusion, and the mycelium will binder all the component together. These components are natural and readily available from agricultural waste, making them interesting as sustainable alternative materials. Indeed, bamboo grows in most regions across the globe, particularly in densely populated areas with limited access to ressources;[37,38] and chitosan is a biopolymer derived from chitin, the major constituent of the shell of shrimps and other crustaceans, and is a food industry waste. When dissolved in a mild acidic aqueous medium, chitosan forms a gel with extrudable viscoelastic properties.[39] First, we select the bamboo fibre size based on the morphology of mycelium grown on them and the mechanical properties of resulting mycelium-bamboo composites. Then, chitosan is added to the mycelium-enriched bamboo fibres to yield a paste that is workable, extrudable and buildable. Finally, we verify the growth of the mycelium on extruded struts and discuss the potential of the method to create mycelium-bound materials with shape complexity. The study and the method proposed could be used by designers, scientists, and engineers to explore the use of mycelium-bound composites for new applications. Further optimisation of the substrate composition could enhance the final properties of the mycelium-bound material.

EXPERIMENTAL SECTION.

**Materials.** The fast growing and commonly available medicinal white-rot fungus, *Ganoderma Lucidum (G. Lucidum),* also known as Lingzhi mushroom, was selected in this study. *G. Lucidum* mycelium was purchased as sawdust spawns from the commercial farm Malaysian Feedmills Farms (Pte) Ltd. Bamboo culms of *Dendrocalamus Asper* were obtained from Indonesia and were processed to sheets in Future Cities

Laboratory (FCL), Singapore, following an established process.[37] Medium molecular weight Chitosan was purchased from Aldrich (200-800 cP viscosity for 1wt% chitosan in 1% acetic acid, acetylation degree 75-85%), Iceland and stirred on a magnetic stirring plate (300 rpm) for 48 hours in an 1% acetic acid solution before use.

**Preparation of bamboo fibres.** The bamboo sheets were first dried in an oven (Qingdao Guosen-1800, China) at 80 °C for a week, after which they were grinded using a cutting mill (Fritsch cutting mill pulverisette 15, Germany). The cut bamboo was then sieved to sizes 1 mm, 500 µm and 200 µm using a vibratory sieve shaker (Fritsch, Analysette 3 Spartan, Germany).

**Preparation of the mycelium-enriched bamboo fibres (enriched fibres).** 175 g of bamboo fibres were hand mixed with 175 mL of deionized water to ensure even distribution of moisture. The pH of the mixture ranged between 6 and 7. The mixture, also called the substrate because it will support the growth of the mycelium, was then packed into a polypropylene bag, autoclaved at 121 °C for 1 hour (Tomy SX-700, Digital Biology, Japan) and left to cool to room temperature. The bag was then opened to mix the sterilized substrate with 50 g of the mycelium-enriched sawdust mother culture, sealed again with a cotton plug and rubber ring, and incubated at 25 °C to 28 °C , relative humidity of 65% to 80% and at pH of 5 to 6 for 1 to 4 weeks in a filtered air environment.[40] The densities of the composite as a function of the mycelium growth were evaluated on mixtures casted in square-shaped moulds and placed on a scale each week.

**Mixing of chitosan with mycelium-enriched bamboo fibres, extrusion and growth.** The mycelium-enriched bamboo fibres were removed from the sealed bags and torn down to smaller pieces. The chitosan solution was added to the enriched fibres at weight ratios of 50%, 60% and 70%, pounded with a pestle and mortar until no more lumps were visible, forming a homogeneous mixture. The pH of the chitosan solution was 6 and was adjusted between 4 and 6 using acetic acid. Extrusion into struts of 5 cm length was carried out manually using serological syringes with 6 mm tip diameter. The mycelium was then left to grow on the extruded struts for 20 days, in air at 23 ± 0.5 °C and 65-70% humidity. **Figure 2** summarizes the processing steps from the mother culture to the extrusion. After 20 days, the samples were dried in an oven (Binder VD 53, Fischer Scientific Pte Ltd, Singapore) and the diameters of the samples were measured using a standard caliper.

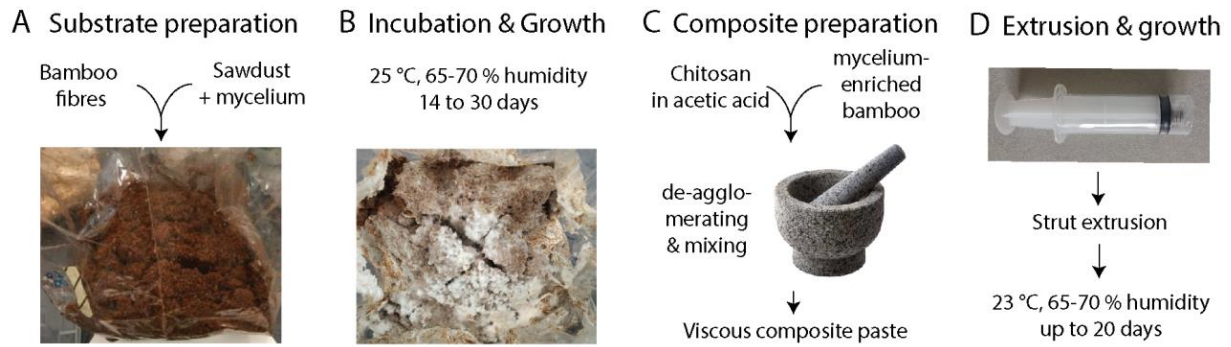

**Figure 2. Flow chart of the fabrication process: (A)** Preparation of the substrate with the bamboo fibres, sawdust and mycelium mother culture, **(B)** growth of the mycelium on the bamboo fibres, **(C)** preparation of the composite paste mixture with de-agglomeration of the mycelium-enriched bamboo fibres using a pestle and mixing with the chitosan solution dissolved in acetic acid, **(D)** extrusion of the composite paste into struts and growth of the mycelium on the struts.

**Microscopy.** Optical microscopic images of the growing network were taken using an upright macroscope (Olympus SZ16 Stereo Zoom Microscope) with the software ImagePro Insight. Mean grey values were determined using the software Image J (NIH, USA) using the blue channel on pictures taken with the same illumination. Electron micrographs were taken with a scanning electron microscope (JSM 5500LV, Jeol, Japan) on carbon-sputtered samples (20 seconds, Pelco SC-6). Prior to sputtering, the samples were dried at 40 °C overnight to stop the mycelium growth and remove moisture. The interconnecting lengths between two branching of the mycelium network were measured using the software Image J and averaged over 50-100 measurements for each image and composition.

**Mechanical testing.** To prepare the samples for tensile testing, the growth of the mycelium was carried out into plastic moulds. The width and length of each sample were of 1 cm and 4 cm, respectively and a thickness of 0.5 cm (see Supporting Information (SI) **Figure S1**). The tensile tests started after 20 days as the samples were too weak to be handled at earlier times. The tests were carried out using a mechanical tester MTS C42 (MTS, USA) using a 50 N load cell, at a strain rate of 0.5 mm/min and using 100 N Bionix grips. The data analysis was done using the TWS Elite software and repeated over four samples in each testing conditions (bamboo fibres size and mycelium growing time). The tangents to the slope of the stress-strain curves were plotted to determine the Young's moduli and the yield stress corresponds to the maximum stress. Compression tests of the chitosan-bamboo-mycelium composites grown after 20 days into square moulds were carried out using an Instron machine (Instron 5565, USA) with a 500 N load cell and a loading rate of 3 mm/min. All tests were repeated in triplicates.

RESULTS AND DISCUSSION.

**Selection of the fibre size for the mycelium-enriched bamboo.**

To prepare mycelium-enriched bamboo fibres to use in the composite paste for to be extruded, mycelium is first grown on the fibres. However, it is known that the mycelium growth kinetics, morphology, and properties are largely influenced by the substrate that provides the physical support and nutrients to the fungus.[41] For example, the oxygen permeation of the substrate is determined by its morphology and packing density and influences how the mycelium develops. Generally, a sufficient oxygen flow is necessary to allow the mycelium to grow homogeneously in bulk and not only at the solid-air interface.[42] To determine which bamboo fibre size to use in our composition, we first incubated the mycelium-enriched sawdust with chopped bamboo fibres of 1 mm, 500 µm and 200 µm-length, and studied the mycelium morphology and properties (**Figure 3**).

All bamboo fibres promoted the growth of the mycelium, with a significant influence of their dimensions on the morphology and density of the mycelium network. This network was denser on smaller bamboo fibre sizes and appeared to grow faster, both in the bulk and in the skin formed at the surface of the substrate exposed to oxygen (**Figure 3A**). The mycelium created an interconnected network with a hypha (the mycelium filaments) of diameter 1.1 ± 0.2 µm. This diameter was constant for all bamboo fibres, but the length between the interconnections in the network was decreasing with growth time and fibre size (**Figure 3B**). The growth of the mycelium in the bulk appeared delayed in the case of the 1 mm fibre length, whereas no significant difference could be measured for the skin after 14 days. After 4 weeks of growth, the skin had an interconnection length of 3.1 ± 0.5 µm for all samples, whereas the bulk had interconnection lengths of 5.7 ± 1.2 µm, 4.5 ± 1.3 µm and 2.9 ± 0.6 µm for 1 mm, 500 µm and 200 µm bamboo fibre length, respectively.

Furthermore, despite the increase in the mycelium network, the overall density of the material decreased with time (**Figure 3C**), indicating that the fungus is consuming the substrate.[43–45] The higher decrease in density for the 200 µm bamboo fibres suggests that smaller fibre sizes promote more mycelium growth (see SI **Figure S2**). Presumably, smaller fibres and larger fibre number increase the surface and reduces the distance between fibres, making the path to the surface shorter and easier for the fungus.

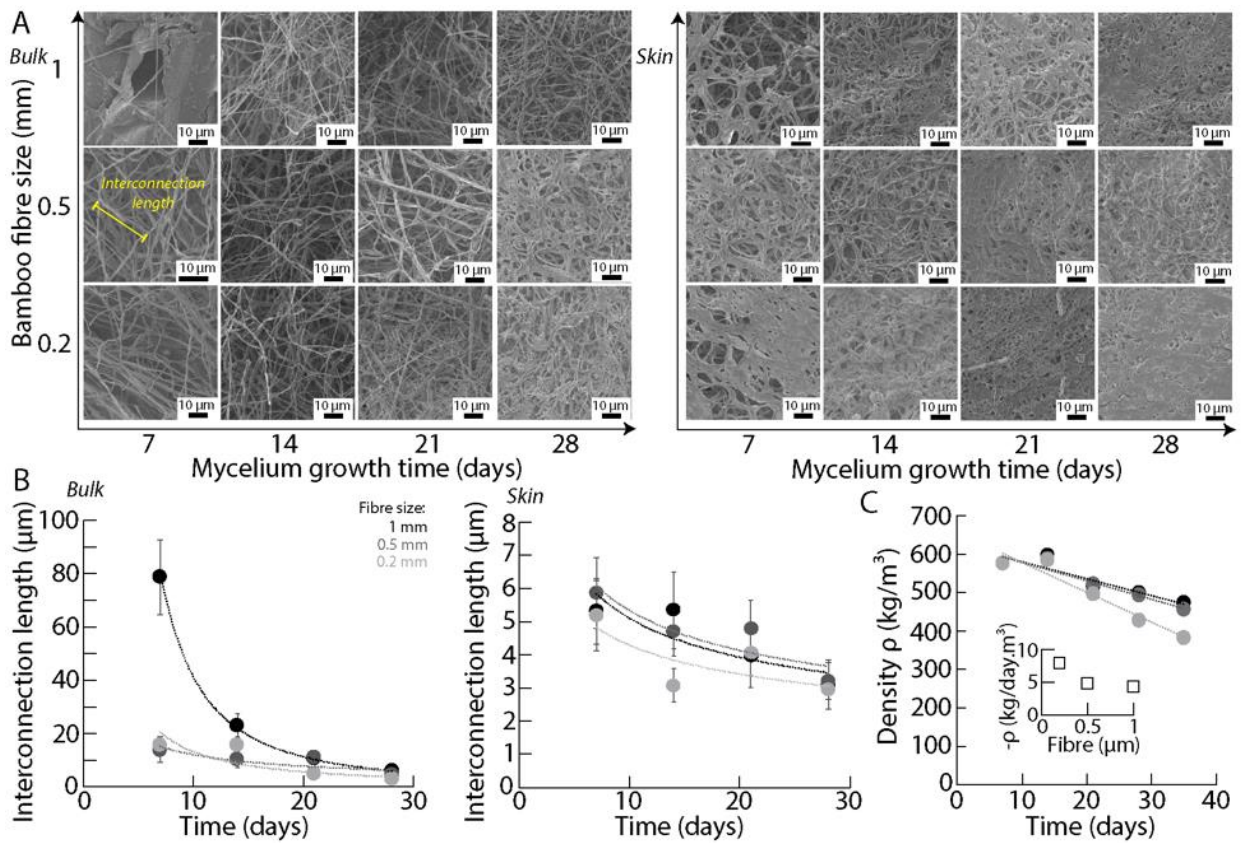

**Figure 3. Microstructure of the mycelium grown on bamboo fibres. (A)** Scanning electron micrographs of the bulk (left) and of the skin (right) of the mycelium grown on bamboo fibres with sizes of 1 mm, 500 µm and 200 µm for 7, 14, 21 and 28 days. **(B)** Interconnection lengths in the mycelium networks grown in the bulk and in the skin as a function of the growing time. **(C)** Density decrease as a function of the time and bamboo fibre size. Insert represents the absolute value of the densification rate as a function of the fibre size. On all graphs, the lines are guides to the eye. Black: 1 mm bamboo fibre; grey: 0.5 mm and light grey: 0.2 mm.

Despite the decrease in density and the degradation of the bamboo fibres by the fungus, the mechanical properties increased as the mycelium has grown (**Figure 4).** The increase in Young's modulus, yield stress and ultimate stress was not significant the 1 mm fibres but was pronounced for 500 and 200 µm. Indeed, the mycelium acts as a binder between the bamboo fibres, providing stress transfer and allowing elongation. When the composite fractured under tension, the hyphae from the mycelium appeared stretched in the tensile direction (see SI **Figure S3**). The mycelium-bamboo with 200 µm fibres exhibited the highest modulus. However, the yield and ultimate stresses were similar between the composites made using the 200 and 500 µm fibres.

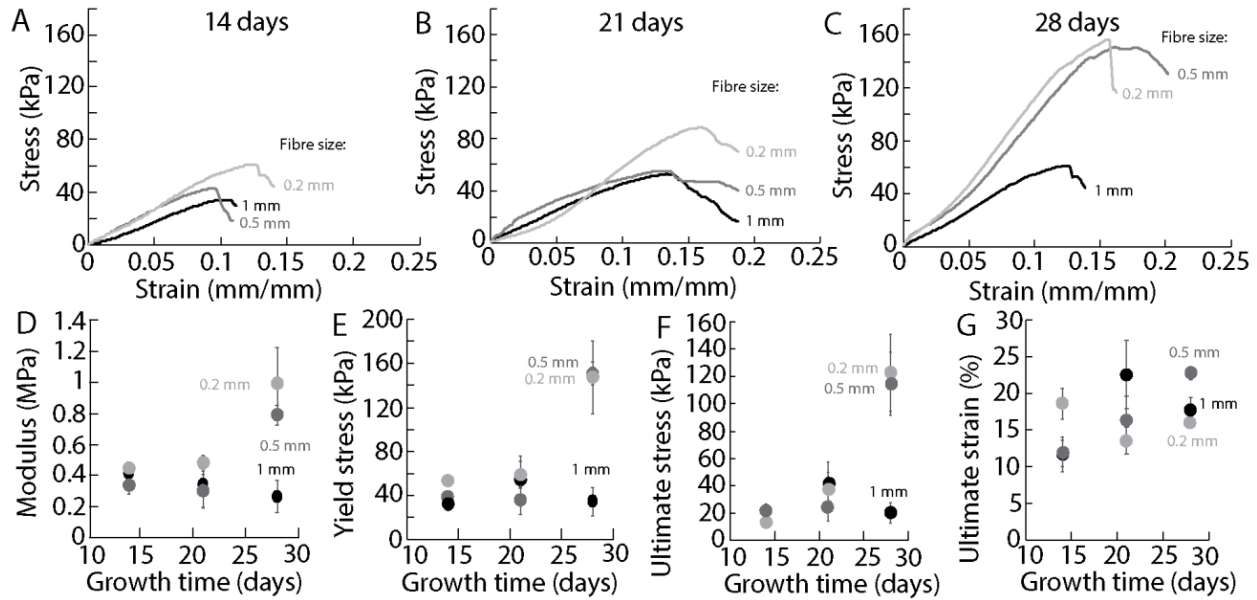

**Figure 4. Mechanical properties of bamboo fibre-mycelium composites. (A-C)** Averaged stress-strain curves of the composites as a function of the bamboo fibre size and mycelium growth time. Modulus **(D)**, yield stress **(E)**, ultimate stress **(F)** and ultimate strain **(G)** under uniaxial tension as functions of the mycelium growth time and bamboo fibre size.

Smaller bamboo fibres sizes are thus more favourable for mycelium growth and stronger mechanical properties. Furthermore, for using a material in a structural application, yield stresses are usually the critical parameters. There was no significant difference between the yield stresses obtained with the 200 and the 500 µm bamboo fibres. Since obtaining chopping and sieving bamboo into 500 µm fibres requires less energy than 200 µm, with comparable mechanical properties, we used the 500 µm fibre size for the mycelium-enriched bamboo.

**Workability and buildability of chitosan-mycelium-enriched bamboo fibre mixtures.**

To process mycelium-bound composites, the common way is to grow the mycelium on a substrate deposited inside a mould.[11,13,16,46] Although moulding is an easy and convenient process, it only allows limited freedom in terms of shape design and complexity. 3D printing methods have been developed to provide shaping and complexity and applied to a variety of materials. [47,48] Direct writing is a 3D printing method that extrudes a paste through a nozzle and therefore requires an ink material that is workable, extrudable, and buildable. To this aim, chitosan dissolved in 1% acetic acid was added as the viscous component (**Figure 5**). Chitosan is an interesting material to work with due to its biodegradability and large availability. In this study, a commercial source of chitosan was used for convenience but we anticipate that similar results could be obtained starting with raw chitin and deacetylating it into chitosan following sustainable processes. [49,50]

We prepared mixtures of chitosan in acetic acid (pH 4) at 2 and 3 wt% with mycelium-enriched bamboo fibres of 500 µm size, and weight ratios chitosan: enriched fibres of 50:50, 60:40 and 70:30. The rationale is to find the optimal composition with: (i) the highest chitosan concentration as possible for good buildability, but with flowing properties. Above 3 wt%, the chitosan solution is a stiff rigid gel and bamboo cannot be added; and (ii) the highest concentration of bamboo fibres to promote mycelium growth and mechanical strength, while maintaining workability and extrudability. The nozzle size used for the extrusion was of 6 mm, and with a horizontal manual set-up. We found from our experiments that all compositions could be extruded in such conditions, although the 70:30 chitosan to fibre ratio showed some exclusion effects with some chitosan solution extruding separately from the mixture, which was not observed for lesser concentrations of chitosan and ratios chitosan: enriched fibres.

Following, we characterized the buildability of the extruded struts by evaluating the evolution of their diameter after 24 hours exposure to air and drying (**Figure 5B**). All struts retained their round shapes and a shrinkage between 3 to 6% was recorded due to the drying. Also, no anisotropic shrinkage was recorded, probably thanks to the random orientation of the bamboo fibres within the struts. A mesh-like extruded structure is shown in **Figure 5C** to demonstrate the buildability of the material.

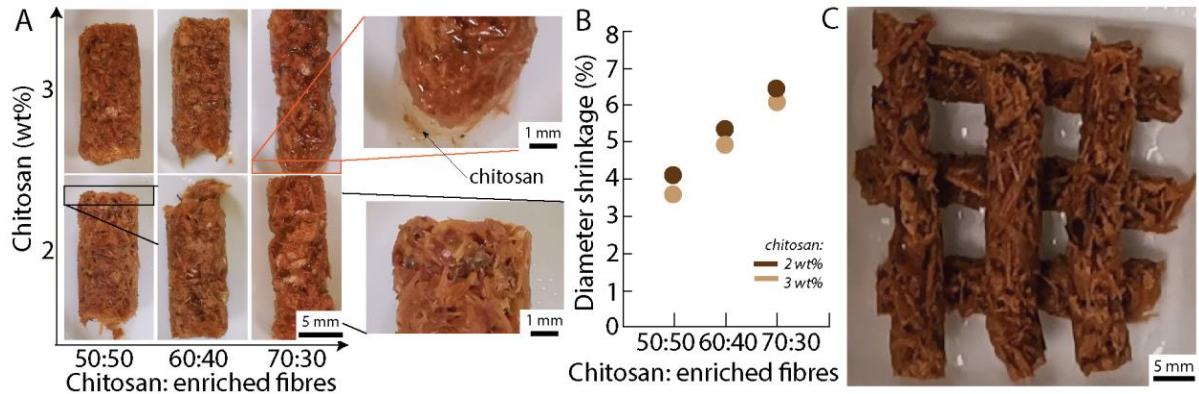

**Figure 5. Extrusion of mycelium-enriched bamboo fibre – chitosan pastes. (A)** Optical images of extruded struts as a function of the chitosan concentration and ratio chitosan to mycelium-enriched bamboo fibres. The magnification is the same for all images in the chart. **(B)** Diameter shrinkage of the struts as a function of the chitosan concentration and chitosan: enriched fibre weight ratio. **(C)** Image of an extruded struts array at 50:50 chitosan: enriched fibre weight ratio.

**Mycelium growth onto extruded struts.**

After extrusion, the mycelium spawn present on the bamboo fibres needs to develop again to consolidate the composite. Despite the antifungal properties of chitosan,[51,52] we observed that mycelium from *G. Lucidum* could grow after drying of the chitosan, in certain pH conditions and ratios chitosan: mycelium-enriched bamboo fibre (**Figure 6**). Indeed, fungi with low-fluidity membranes and high glucan-chitin ratios in their cell walls are resistant to chitosan,[53] and *G. Lucidum* is one of them.[54–56]

After 20 days in air at temperature of ~25 °C and 65-70% humidity, the mycelium grew again on the extruded struts (**Figure 6A**). This was visualised optically with a white appearance, contrasting with the orange-brown colour of the initial bamboo-chitosan substrates. Despite a non-sterilized environment, only 1/5th of the struts showed mild contamination, denoting the benefit of using antibacterial chitosan as a matrix. Furthermore, more mycelium had grown on the struts with a high concentration of chitosan of 3 wt% as compared to 2 wt%, and with a high ratio of chitosan to enriched bamboo fibres (**Figure 6B,C**).

This could be due to the fact that a higher concentration of chitosan dries faster and brings the bamboo fibres close together. Indeed, it was observed that mycelium started to grow as soon as most of the water and acetic acid had evaporated. In addition, the pH of the mixture had a significant effect, with a pH of 6 showing the highest mycelium growth. Although the effect of pH on the growth of *G. Lucidum* is not clear,[57–59] it is likely that pH 6 in our composition is favoured because it is close to the iso electric point of chitosan. The morphology of the mycelium grown at pH 6 was different from the mycelium grown on the mycelium-enriched bamboo without chitosan and appeared sprinkled with hollow tubes and intensive branching (**Figure 6D,E**). Highly branched morphologies were observed in other species grown in presence of chitosan.[60,61] The width of the hollow mycelium fibres was of $0.85 \pm 0.15$ µm. These experiments thus show that mycelium can grow again onto the extruded struts to bind and consolidate the extruded structures, converting the paste into a stiff material, and this despite the antimicrobial properties of chitosan.

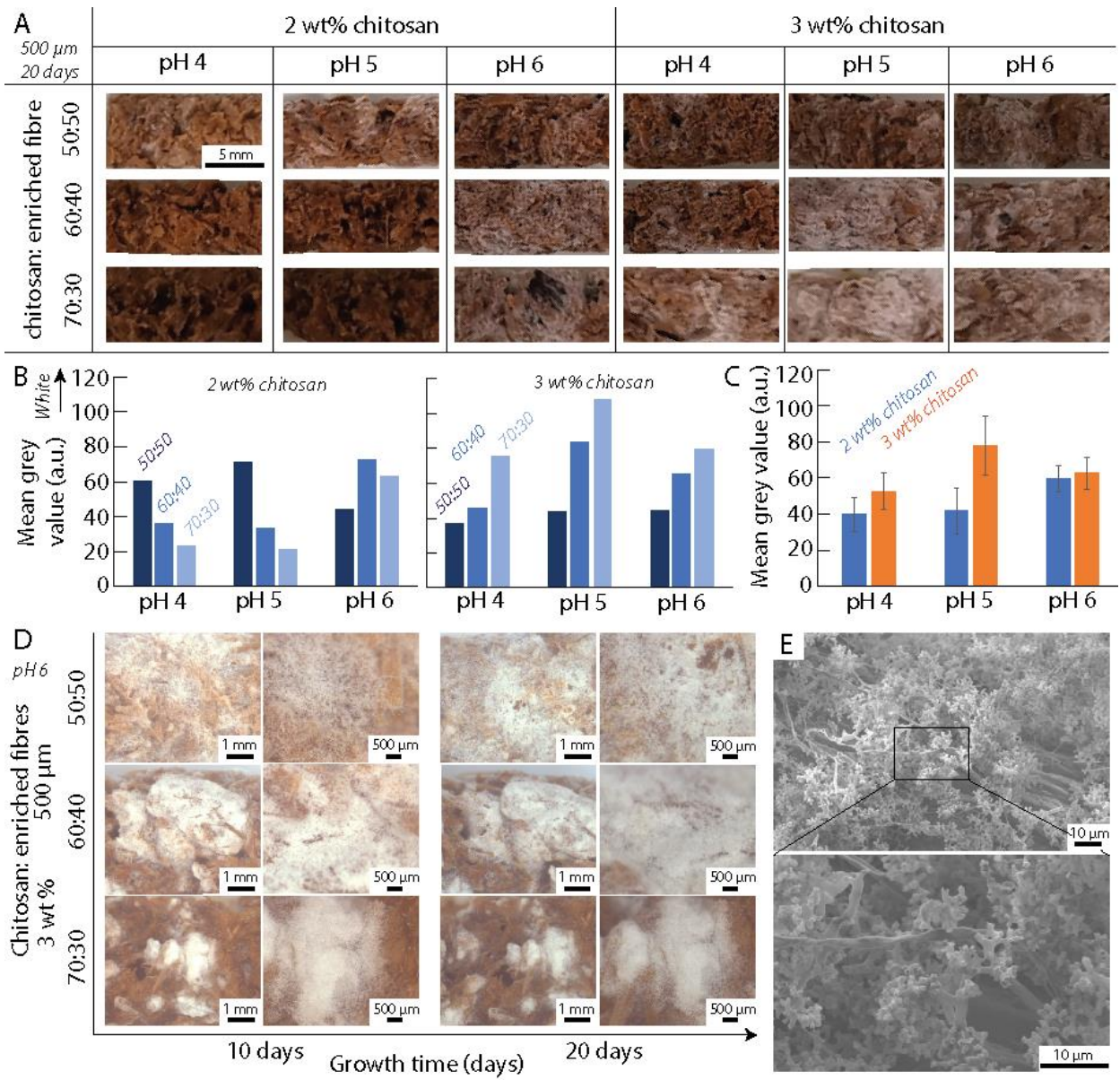

**Figure 6. Mycelium growth on extruded struts.** **(A)** Optical images of the struts made with 500 μm mycelium-enriched bamboo fibres, for chitosan concentrations of 2 and 3 wt%, pH of 4 to 6 and chitosan: enriched fibres weight ratio of 50:50, 60:40 and 70:30. The scale is the same for all pictures. **(B)** Mean grey values of the images as a function of pH, chitosan concentration and chitosan: enriched fibres weight ratio. A high mean grey value indicates a high growth of mycelium. **(C)** Grey values summarized on all samples as a function of pH and chitosan concentration. **(D)** Images of the struts extruded at pH 6, 3wt% chitosan and 500 μm mycelium-enriched bamboo fibre as a function of the growth time and chitosan: enriched fibre

ratio. The white areas are the mycelium. **(E)** Electron micrographs showing the morphology of the mycelium forming in the struts.

CONCLUSIONS.

The mixture of chitosan with mycelium-enriched bamboo developed in this study is promising for building mycelium-bound composites with complex 3D shapes. The compositions that work best to combine workability, extrudability and buildability used 500 µm mycelium-enriched bamboo fibres mixed at a chitosan: enriched fibres ratio of 60:40 or 70:30 and with a 3wt% chitosan solution prepared at pH~6. Furthermore, the use of chitosan did not compromise the growth of the mycelium. The composition, tested under compression after 20 days of mycelium growth, yielded a compression modulus of 40 kPa, compared to 240 kPa without chitosan (see **Figure S4** for the compression curves) (**Figure 7**). Despite the decrease in mechanical properties, the composition has the extra degree of capability of extrudability for complex shaping, which are requirements for direct writing of structures and the further exploration of mycelium-bound materials for structural applications. For example, automation of the printing process, printing of struts geometries with various patterns and their influence on the growth and the properties and optimisation of the substrate composition using other additives are anticipated future directions of research.

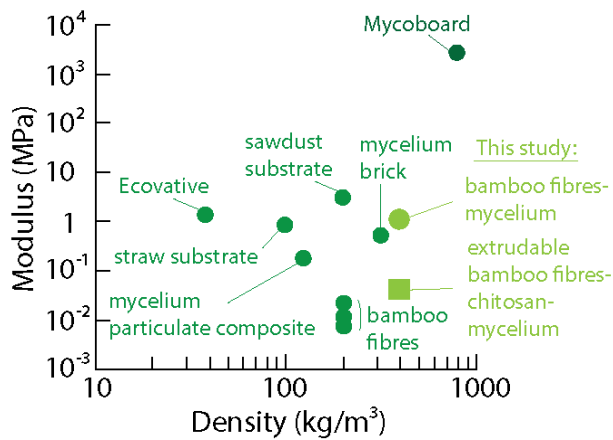

**Figure 7.** Modulus as a function of the density for mycelium-bound materials. Data extracted from this study and other references.[2,6,62–64]

HIGHLIGHTS

- Mycelium-bound composites are sustainable alternative to current composites used in architecture, for example.
- Having an extrudable composition to build mycelium-bound composites allows complex shaping and designing.
- Addition of chitosan to mycelium-enriched bamboo fibres provides the workability, extrudability, and buildability required while allowing mycelium to grow.

ASSOCIATED CONTENT. **Supporting Information**.

A PDF file is included as supporting information, comprising the following figures:
**Figure S1:** Tensile test of the bamboo fibres-mycelium composites.

**Figure S2:** Growth of the mycelium onto the bamboo fibres.

**Figure S3:** Stretching of the hyphae network under tension.

**Figure S4:** Compression curves of the bamboo fibres-mycelium and bamboo fibres-chitosan-mycelium composites.

AUTHORS INFORMATION

**Corresponding Author**

* hortense@ntu.edu.sg

**Present Addresses**

†Nanyang Technological University, School of Materials Science and Engineering, 50, Faculty Avenue, Singapore 639798.

§Alternative Construction Materials, Future Cities Laboratory, Singapore-ETH Centre, 1 Create Way, Singapore 138602.

+Sustainable Construction, Faculty of Architecture, Karlsruhe Institute of Technology, Englestrasse 11, Karlsruhe, Germany 76131.

‖Nanyang Technological University, School of Mechanical and Aerospace Engineering, 50, Faculty Avenue, Singapore 639798.

**Author Contributions**

The manuscript was written through contributions of all authors. All authors have given approval to the final version of the manuscript. ‡These authors contributed equally.

**ORCID**

Hortense Le Ferrand 0000-0003-3017-9403

**Notes**

The authors declare no competing financial interest.

**Data availability**

The authors confirm that the data supporting the findings of this study are available within the article and its supplementary materials, or upon request to the corresponding author.

ACKNOWLEDGMENT

The authors thank the microscopy centre (Facilities for Analysis, Characterisation, Testing and Simulations) from Nanyang Technological University, Singapore. The authors acknowledge start up fund from Nanyang Technological University, Singapore.

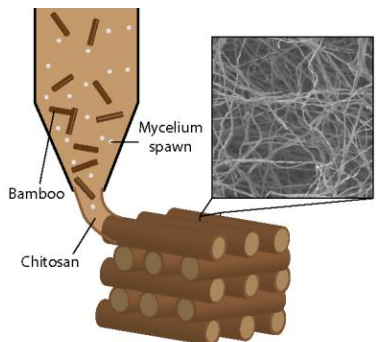

TOC.

SUPPORTING INFORMATION.

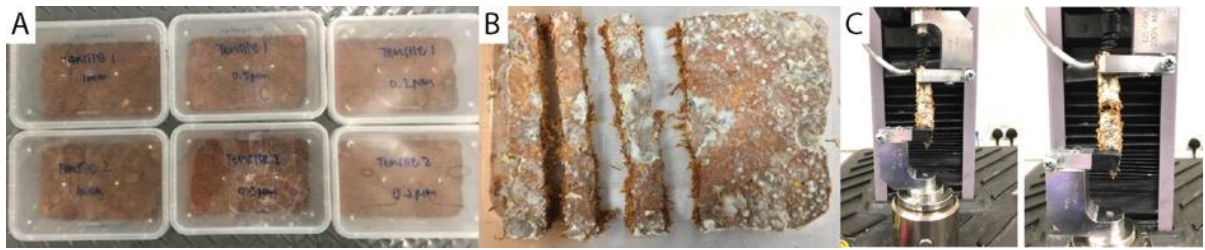

**Figure S1: Tensile test of the bamboo fibres-mycelium composites. (A)** Sample preparation in square moulds of 4 cm width, **(B)** Cutting stripes of width 1 cm, **(C)** Tensile testing of the specimens.

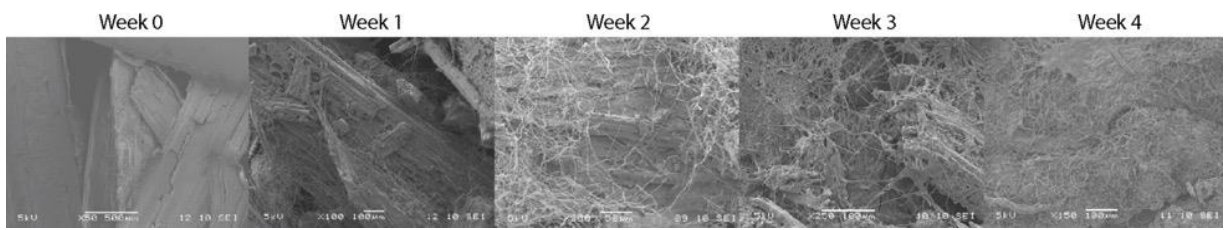

**Figure S2: Growth of the mycelium onto the bamboo fibres.** Electron micrographs showing the growth of the mycelium on the 200 µm bamboo fibres from 0 to 4 weeks growth.

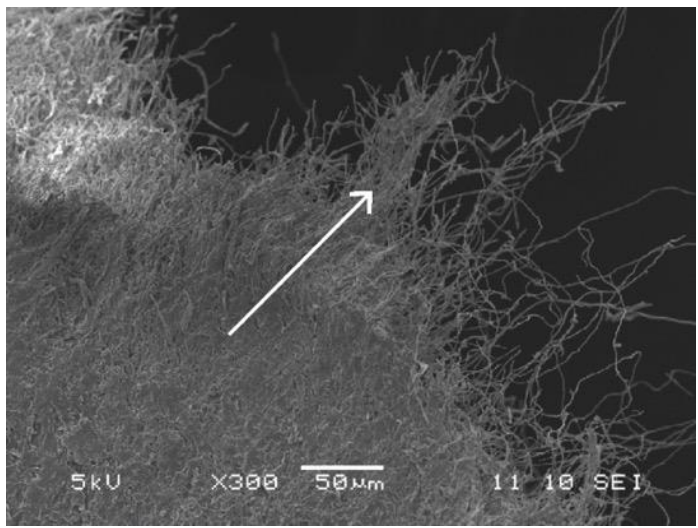

**Figure S3: Stretching of the hyphae network under tension.** Electron micrograph of the stretched mycelium grown for 3 weeks onto the 200 µm-bamboo fibres substrate. The arrow indicates the tensile direction.

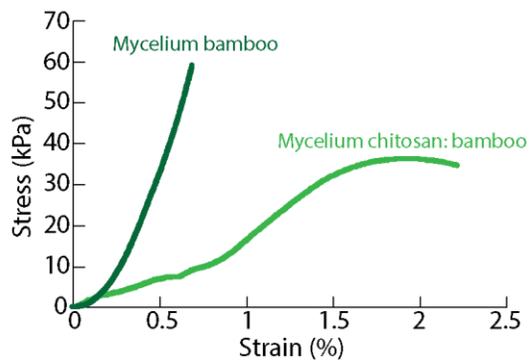

**Figure S4: Representative compression curves of the bamboo fibres-mycelium and bamboo fibres-chitosan-mycelium composites.**